\documentclass{article}
\usepackage{jheppub}
\usepackage{graphicx}
\usepackage{dcolumn}
\usepackage{bm}
\usepackage{physics}
\usepackage{amsmath,amsfonts,amssymb}
\usepackage{subcaption}
\usepackage{setspace}
\usepackage{color}
\usepackage{tikz}
\usepackage{float}
\usepackage{hyperref}
\usepackage{parskip}
\usepackage{cleveref}
\usepackage[normalem]{ulem}
\allowdisplaybreaks

\title{Momentum Space Correlation Functions in 2D Galilean Conformal Algebra }

\author[a]{Anchita Chetia}
\author[a]{Nirmalya Kajuri}
\author[b]{Chandra Prakash
\footnote{Author names are listed alphabetically, following standard practice in high-energy theory.}}

\affiliation[a]{School of Physical Sciences, IIT Mandi, Himachal Pradesh, India}
\affiliation[b]{Department of Physics, Indian Institute of Technology, Guwahati, 
Assam, India}

\emailAdd{di2405@students.iitmandi.ac.in}
\emailAdd{nirmalya@iitmandi.ac.in}
\emailAdd{chandra.pp@alumni.iitg.ac.in}

\abstract{Galilean Conformal Algebra (GCA) arises as a controlled nonrelativistic limit of the relativistic conformal algebra. In this paper, we initiate the study of momentum space correlation functions in two-dimensional GCA. We derive and solve momentum space Ward identities to obtain two-point and three-point functions. However, relating them to position space correlation functions presents a challenge as Fourier transforms of the latter do not exist. This is resolved by analytically continuing the boost eigenvalues to imaginary values. In this regime, the Fourier transform of the position space two-point and three-point functions exist and match exactly with the momentum space two-point and three-point function obtained by solving the Ward identities.}

\date{} 

\begin{document}

\maketitle
\section{Introduction}
The Galilean Conformal Algebra (GCA) arises as a contraction of the relativistic conformal algebra in the limit of infinite speed of light. Its role relative to conformal algebra is analogous to that of the Galilean algebra relative to the Poincaré algebra: it captures the residual symmetry structure that survives a controlled non-relativistic limit. There is also an intrinsic characterization of GCA--it is the maximal extension of the Galilean algebra that incorporates scale invariance and special conformal transformations. 

GCA appears in several physical contexts. It was first introduced as a non-relativistic limit of AdS/CFT \cite{Bagchi:2009my}, where boundary conformal symmetry contracts to a Galilean conformal symmetry while the dual bulk dynamics approaches a Newton–Cartan or related non-Lorentzian geometry \cite{Alishahiha:2009np,Martelli:2009uc,Duval:2009vt,Sakaguchi:2009de,Hotta:2010qi,Bagchi:2010vw,Duval:2011mi,Fedoruk:2011hi,Bagchi:2013qva,Rasmussen:2017eus,Bergshoeff:2017btm,Malvimat:2018izs,Lovrekovic:2022lwv,Hartong:2024lqb}. Closely related is the appearance of the GCA in flat-space holography, where the asymptotic symmetry group of three-dimensional flat spacetime—the BMS$_3$ algebra—is isomorphic to the infinite-dimensional two-dimensional GCA \cite{Bagchi:2010zz,Bagchi:2012cy,Bagchi:2012yk,Bagchi:2012xr,Bagchi:2013bga,Bagchi:2014iea,Basak:2022cjs,Basak:2022gcv}. This BMS/GCA correspondence suggests that GCA may play a role in formulating holography beyond AdS. Aspects of GCA and its extensions have been studied extensively in the literature \cite{deAzcarraga:2009ch,Bagchi:2009ke,Mukhopadhyay:2009db,Hosseiny:2009jj,Mandal:2010gx,Bagchi:2010xw,Fedoruk:2011ua,Setare:2011hc,Hosseiny:2011ct,Bagchi:2014ysa,Basu:2015evh,Mandal:2016lsa,Mandal:2016wrw,Batlle:2016iel,Bagchi:2015qcw,Casali:2017zkz,Bagchi:2017yvj,Barducci:2018wuj,Lodato:2018gyp,Chen:2019hbj,Banerjee:2019axy,Ragoucy:2020elp,Bagchi:2021qfe,Mehra:2021sfx,Campoleoni:2021blr,Ragoucy:2021iyo,Basu:2021axf,Basu:2021awn,Bagchi:2022twx,Basu:2022nyl,Fontanella:2024hgv}, including a position space bootstrap program \cite{Bagchi:2009ca,Bagchi:2009pe,Setare:2010nf,Chen:2020vvn,Chen:2021xkw,Chen:2022cpx}.

Existing work on GCA-invariant field theories has focused solely on position space. Meanwhile, momentum-space methods have emerged as a powerful framework for conformal field theories \cite{Bzowski:2013sza,Coriano:2013jba,Bzowski:2015pba,Isono:2018rrb,Bzowski:2019kwd,Gillioz:2019lgs,Gillioz:2021sce,Jain:2020rmw,Jain:2021wyn,Marotta:2022jrp,Cardona:2018dov,Bautista:2019qxj,Bzowski:2020kfw,Gillioz:2025yfb}. It has also been studied for the ultrarelativistic Carroll group \cite{Marotta:2025qjh} and the nonrelativistic Schr\"odinger group \cite{Mehen:1999nd,Gupta:2024yib}.

In this approach, conformal symmetry is imposed on correlation functions via momentum space Ward identities. In relativistic CFTs, momentum-space correlators offer a complementary perspective to position space. They make manifest the analytic properties, factorization channels, and singularity structure that are often obscured in coordinate space. Apart from conformal bootstrap, momentum space methods have proven useful in several areas such constraining cosmological correlators \cite{McFadden:2009fg,Maldacena:2011nz,Mata:2012bx,Bzowski:2012ih,Assassi:2012zq,Creminelli:2012ed,Hinterbichler:2013dpa,Kundu:2014gxa,Ghosh:2014kba,Arkani-Hamed:2018kmz,Baumann:2019oyu,Sleight:2019hfp,Pajer:2020wxk,Baumann:2020dch,Goodhew:2020hob,Melville:2021lst,DiPietro:2021sjt,Hogervorst:2021uvp,Baumann:2022jpr,Jain:2022uja,Jazayeri:2022kjy,Pimentel:2022fsc,Bissi:2023bhv,Dey:2025kci} , and relating correlators with scattering amplitudes \cite{Raju:2011mp,Marotta:2024sce,Raju:2012zs,Farrow:2018yni,Hijano:2019qmi,Hijano:2020szl,Gillioz:2020mdd,Albayrak:2020isk,Jain:2021qcl,Gadde:2022ghy,Prabhu:2024khb}.

In this work, we initiate a systematic study of momentum-space correlation functions for the 2-dimensional Galilean conformal algebra. We focus on scalar primaries. Our work offers both a fresh perspective on GCA and a new testing ground for momentum space methods.  

The reason for the focus on two dimensions is twofold. In 2D, the nonrelativistic contraction maps the pair of Virasoro algebras to an infinite-dimensional GCA. This is in contrast to higher dimensions, where it yields a finite-dimensional algebra. GCA$_2$ thus provides a sharply constrained setting and the one directly relevant to the BMS$_3$. Secondly, it is in 2D that we have the exponentially growing position-space correlators, which presents an interesting challenge for momentum space methods. In higher dimensions, the problem simplifies as the position-space correlators exhibit power law dependence. 

We find momentum space GCA$_2$ two-point and three-point functions by solving the Ward identities and obtain well-defined results. However, in two dimensions, position-space GCA correlators exhibit exponential dependence on spatial separations. They are not tempered distributions, and their Fourier transform does not exist for real momenta. Thus, while both the position and momentum space correlators exist, they cannot be related via a Fourier transform. 

To relate the two, we analytically continue the boost eigenvalues to be imaginary\footnote{An equivalent alternate route is to analytically continue the Fourier-transformed correlation functions from imaginary to real momentum.}. This takes us to a regime where both sets of correlators are tempered distributions. The Fourier transform of position space two-point and three-point functions is then well-defined and matches exactly with the result obtained by solving momentum-space Ward identities. As an independent check for the two-point function, we arrive at the same answer by taking the appropriate nonrelativistic limit of the relativistic conformal two-point function.

The paper is organized as follows. We start with a review of GCA in position space. In section II, we obtain the momentum space Ward identities for GCA. Section III presents our computation for the two-point function. As mentioned earlier, we obtain it in three different ways-- Fourier transforming the position space GCA two-point function, solving the Ward identities and the nonrelativistic limit of the relativistic conformal two-point function. We turn to the three-point function in section IV and show that the answers from the Fourier transform and direct solution match. We conclude the paper with a summary and future plans.

\section{Review of position space GCA}
We start by providing a brief review of position-space GCA. For more details, we refer the reader to \cite{Bagchi:2009ca,Chen:2020vvn}. 
\subsection{GCA from Wigner-In\"on\"u contraction of conformal algebra}

We begin with the conformal algebra in $(d+1)$-dimensional Minkowski spacetime, generated by translations $P_\mu$, Lorentz transformations $J_{\mu\nu}$, dilatations $D$, and special conformal transformations $K_\mu$, with $\mu = 0,1,\ldots,d$. The Galilean Conformal Algebra (GCA) is obtained as a non-relativistic contraction of this algebra.

The contraction corresponds to a limit in which spatial coordinates scale relative to time. At the level of spacetime coordinates, this is implemented by
\[
x^i \rightarrow \epsilon\, x^i, 
\qquad 
t \rightarrow t,
\]
with $\epsilon \to 0$. Equivalently, this may be viewed as sending the speed of light $c \to \infty$. The contraction is defined entirely at the level of the boundary Minkowski spacetime.

The Lorentz generators decompose into spatial rotations $J_{ij}$ and boosts $J_{0i}$. Under the above scaling, the rotations remain unscaled,
\[
J_{ij} = x_i \partial_j - x_j \partial_i,
\]
while the boosts must be rescaled to remain finite. Defining
\[
B_i \equiv \epsilon\, J_{0i},
\]
and using $J_{0i} = x_0 \partial_i - x_i \partial_0$, one finds
\[
B_i = \lim_{\epsilon \to 0} \epsilon \left( t\, \partial_i - \epsilon x_i \partial_t \right)
= t\, \partial_i .
\]

The translation generators split into time and space translations,
\[
H \equiv P_0 = \partial_t,
\qquad
P_i = \partial_i,
\]
which remain finite in the $\epsilon \to 0$ limit.

The relativistic dilatation generator
\[
D = -x^\mu \partial_\mu
\]
contracts straightforwardly to
\[
D = -\left( t\, \partial_t + x^i \partial_i \right),
\]
which generates simultaneous scaling of space and time.

The special conformal generators are given by
\[
K_\mu = 2 x_\mu (x \cdot \partial) - (x \cdot x)\, \partial_\mu .
\]
Separating temporal and spatial components and applying the non-relativistic scaling, the temporal component contracts to
\[
K \equiv K_0
= -\left( t^2 \partial_t + 2t\, x^i \partial_i \right),
\]
while the spatial components contract to
\[
K_i = t^2 \partial_i .
\]

The resulting finite-dimensional Galilean Conformal Algebra in $d$ spatial dimensions is generated by
\[
\left\{ J_{ij},\, P_i,\, H,\, B_i,\, D,\, K,\, K_i \right\},
\]
where $J_{ij}$ generate spatial rotations, $P_i$ generate spatial translations, $H$ generates time translations, $B_i$ generate Galilean boosts, $D$ generates dilatations, $K$ generates temporal special conformal transformations, and $K_i$ generate spatial special conformal transformations.

\subsection{The Galilean Conformal Algebra}
The algebra of the generators obtained above forms the Galilean Conformal algebra (GCA). The non-zero commutators are:

\begin{align*}
[J^{ij}, J^{rs}] &= \delta_{ir} J^{js} - \delta_{is} J^{jr}
                  - \delta_{jr} J^{is} + \delta_{js} J^{ir}, \\[4pt]
[J^{ij}, B^r] &= B^i \delta_{jr} - B^j \delta_{ir}, &
[J^{ij}, P^r] &= P^i \delta_{jr} - P^j \delta_{ir}, &
[J^{ij}, K^r] &= K^i \delta_{jr} - K^j \delta_{ir}, \\[6pt]
[H, B^i] &= - P^i, &
[K, B^i] &= K^i, &
[K, P^i] &= 2 B^i,  \\[6pt]
[D, H] &= H, &
[D, K^i] &= -K^i, &
[D, P^i] &= P^i, \\[4pt]
[H, K] &= -2 D, &
[D, K] &= -K ,&
[H, K^i] &= -2 B^i.
\end{align*}
We introduce the notation that we will use henceforth:
\[
\begin{aligned}
&L_{-1} = H, \qquad
L_{0} = D, \qquad
L_{1} = K, \\[6pt]
&M^i_{-1} = P^i, \qquad
M^i_{0} = B^i, \qquad
M^i_{1} = K^i .
\end{aligned}
\]
The form of the algebra becomes:
\[
\begin{aligned}
[J^{ij}, L_{n}] &= 0, \\
[L_{m}, M^i_{n}] &= (m-n)\, M^i_{(m+n)}, \\
[J^{ij}, M^k_{m}] &= M^i_{m} \delta_{jk} - M^j_{m} \delta_{ik}, \\
[M^i_{m}, M^j_{n}] &= 0, \\
[L_{m}, L_{n}] &= (m-n)\, L_{m+n} .
\end{aligned}
\]
In the 2D case, $J^{ij}$ drop out. GCA$_2$ is generated by $L_m,M^i_n$.

\subsection{Representation and correlation functions}\label{cor}

The representation of GCA is obtained in the usual way by first considering the subalgebra that leaves the origin invariant:
\begin{align}
& {\left[L_0, O(0)\right]=\Delta O(0)} \\
& {\left[M_0^i, O(0)\right]=\xi^i O(0)}
\end{align}
Here $\Delta, \xi_i$ are the eigenvalues of dilatation and Galilean boost, respectively. 

The action of the generators on a primary at an arbitrary point can be derived from the subalgebra and is given by: 
\begin{align}\label{l}
& {\left[L_n, O(x, t)\right] }=t^{n+1} \partial_t+(n+1) t^n x^i \partial_i+(n+1)\left(t^n \Delta -n t^{n-1} x_i \xi^i\right) O(x, t) \\ \label{m}
& {\left[M_n^i, O(x, t)\right] }=\left[-t^{n+1} \partial_i+(n+1) t^n \xi^i\right] O(x, t)
\end{align}

From \eqref{l} and \eqref{m}, the Ward identities for two-point and three-point functions follow. The Ward identities fix the position-space two-point and three-point function completely, as was first shown in \cite{Bagchi:2009ca}. For 2D GCA, the two-point function is given by:
\begin{align}
\label{2p}
G^{(2)}(x,\tau)
=C^{(2)}\,
\delta_{\Delta_1,\Delta_2}\,
\delta_{\xi_1,\xi_2}\,
\tau^{-2\Delta_1}
\exp\!\left(\frac{2\,\xi_1\, x}{\tau}\right)
\end{align}
where $C^{(2)}$ is an arbitrary constant.
The three-point function has the form:
\begin{align}
G^{(3)}= C_{123}\,
|t_{12}|^{-\Delta_{12,3}}
|t_{23}|^{-\Delta_{23,1}}
|t_{31}|^{-\Delta_{31,2}} \,
\exp\!\left[
\left(
\xi_{12,3} k_{12}
+ \xi_{23,1} k_{23}
+ \xi_{31,2} k_{31}
\right)
\right].
\end{align}
where
\begin{align}
\Delta_{ij,k} &= \Delta_i + \Delta_j - \Delta_k,
\qquad
\xi_{ij,k} = \xi_i + \xi_j - \xi_k, \\[6pt]
k_{ij} &= \frac{x_{ij}}{t_{ij}} .
\end{align}

With no loss of generality, we use translational invariance to set
\[
x_3 = 0, \qquad t_3 = 0 .
\]
The three-point function then takes the form:
\begin{align}
G^{(3)}=|t_1 - t_2|^{-\Delta_{12,3}}
|t_2|^{-\Delta_{23,1}}
|t_1|^{-\Delta_{31,2}}
\exp\!\left[\xi_{12,3} \frac{x_1 - x_2}{t_1 - t_2}
+ \xi_{23,1} \frac{x_2}{t_2}
+ \xi_{31,2} \frac{x_1}{t_1}\right].
\end{align}

Both the two-point and three-point functions diverge with distance, which implies that their Fourier transform does not exist. As we will see, continuing the boost eigenvalues to imaginary values allows us to take the Fourier transform and compare with momentum space results. For later convenience, we write down the two-point and three-point functions with $\xi \to -i \xi$ (the minus sign is a matter of convention):
\begin{align}
\label{2pi}
G^{(2)}(x,\tau)
=C^{(2)}\,
\delta_{\Delta_1,\Delta_2}\,
\delta_{\xi_1,\xi_2}\,
\tau^{-2\Delta_1}
\exp\!\left(\frac{-2\,i\xi_1\, x}{\tau}\right)
\end{align}
\begin{align}
\label{3p}
G^{(3)}=|t_1 - t_2|^{-\Delta_{12,3}}
|t_2|^{-\Delta_{23,1}}
|t_1|^{-\Delta_{31,2}}
\exp\!\left[-i\left(\xi_{12,3} \frac{x_1 - x_2}{t_1 - t_2}
+ \xi_{23,1} \frac{x_2}{t_2}
+ \xi_{31,2} \frac{x_1}{t_1}\right)\right].
\end{align}
\subsection{Correlation functions from nonrelativistic limit}\label{psn}
It was shown in \cite{Setare:2010nf} that the GCA two-point function can also be obtained by taking a nonrelativistic limit of the relativistic conformal two-point function. Here we recall this limit in some detail, as we will use it to check momentum space two-point functions as well. 

Consider a 2D Lorentzian CFT. Denote $u=c(\tau+\gamma x), v=c(\tau-\epsilon x)\ $ where $\tau=t_{12},x=x_{12}$ and $\gamma=1/c.$. 
We have the operators
\[
L_0 = - u \partial_u, \qquad
\bar L_0 = - v \partial_{v},
\]
with eigenvalues\( h, \bar{h} \) respectively. The dilatation operator is \(D = L_0 + \bar{L}_0,\) with dilatation eigenvalue
\(\Delta = h + \bar{h}\).

The Galilean boost is obtained in the limit:
\[
- t \partial_x
= \lim_{c \to \infty} \frac{1}{c}\,(L_0 - \bar{L}_0).
\]

If the nonrelativistic limit is taken simply as $c\to \infty$, the Galilean boost eigenvalues would vanish:
\[
\lim_{c \to \infty} \frac{1}{c}(h - \bar{h}) = 0 .
\]
Following \cite{Setare:2010nf}, we choose
\[
h = \frac{\Delta}{2} + \frac{\xi}{2\gamma},
\qquad
\bar{h} = \frac{\Delta}{2} - \frac{\xi}{2\gamma},
\]

Then in the $c \to \infty$ limit,
\[
\frac{1}{c}(h - \bar h) \sim 2  \xi .
\]
The Wightman two-point function in 2d CFT scales as
\[G^{(2)}\propto
(u-i\varepsilon)^{-2h}\,(v-i\varepsilon)^{-2\bar{h}}.
\]
In what follows, we suppress the $i\varepsilon$ for simplicity. 
Substituting the scaling:
\begin{align}
\label{nr1}
&= (\tau+\gamma x)^{-\left(\frac{\Delta}{2}+\frac{\xi}{2\gamma}\right)}
(\tau-\gamma x)^{-\left(\frac{\Delta}{2}-\frac{\xi}{2\gamma}\right)} .
\end{align}

Rewriting
\begin{align}
\label{nr2}
(\tau \pm \gamma x)^{\pm \xi/{2\gamma}}
= \exp\!\left[\pm \frac{\xi}{2\gamma}\ln(\tau \pm \gamma x)\right].
\end{align}
and Taylor expanding
\begin{align}
\label{nr3}
\ln(\tau \pm \gamma x)
= \ln \tau \pm \frac{\gamma x}{\tau} +\mathcal{O}(\gamma^2)
\end{align}

Substituting \eqref{nr2} and \eqref{nr3} in \eqref{nr1} and taking the $\gamma\to 0$ limit, we get 
\[G^{(2)}(t_{12},x_{12})=
\tau^{-\Delta}\,
\exp\!\left(\frac{2\xi x}{\tau}\right)=t_{12}^{-\Delta}\,
\exp\!\left(\frac{2\xi x_{12}}{t_{12}}\right).
\]
Which matches \eqref{2p}. If we took $\xi \to -i \xi$, we would recover \eqref{2pi}. 

\section{Momentum space representation of 2D GCA}
In this section, we derive the representation of GCA in momentum space, starting from the position space representation. We will work in general dimensions in this section.

We start with the commutation relation:
\begin{align}
\label{vectorward}
[M_n^i,\, O(x,t)] = \big[-t^{n+1}\partial_i + (n+1)t^n \xi^i\big]\,O(x,t)
\end{align}

To obtain its momentum space counterpart, we introduce the momentum space primary $\widetilde{O}$:
\begin{align}
O(x,t)
&= \int \frac{dE\, d^d k}{(2\pi)^{d+1}}\;
e^{-iE t + i\mathbf{k}\cdot\mathbf{x}}\,
\widetilde{O}(E,\mathbf{k}) \\
\text{where } \,
\widetilde{O}(E,\mathbf{k})
&= \int dt \, d^d x\;
e^{iE t - i\mathbf{k}\cdot\mathbf{x}}\, O(x,t)
\end{align}

Now we Fourier transform both sides of \eqref{vectorward} and obtain:
\[[M_n^i,\, \widetilde{O}(E,\mathbf{k})]=- \int dt\,d^dx\, e^{iE t - i\mathbf{k}\cdot\mathbf{x}}\, \left(t^{n+1}\partial_i O(x,t)+(n+1)\xi^i t^n O(x,t)\right)\]

which simplifies to:
\begin{align}
\label{vectorwardmo}
[M_n^i,\, \widetilde{O}(E,\mathbf{k})]
= (-i)^n\Big[-k_i \partial_E^{\,n+1}
+ (n+1)\xi^i \partial_E^{\,n}\Big]\widetilde{O}(E,\mathbf{k}).
\end{align}

Next, we consider the commutator:
\[
[L_n,\, O(x,t)]
= \big[t^{n+1}\partial_t
+ (n+1)t^n x^i\partial_i
+ (n+1)(t^n\Delta - n t^{n-1} x_i \xi^i)\big]\,O(x,t).
\]

Following the same steps as in the previous case, we obtain: 
\begin{align}
\notag
[L_n,\widetilde{O}(E,\mathbf{k})]
= (-i)^n\Big[
- E\partial_E^{\,n+1}\widetilde{O}(E,\mathbf{k})
&+ (n+1)(-k_i\partial_{k_i}-d-1+\Delta)\partial_E^{\,n}\widetilde{O}(E,\mathbf{k})\\
&+ (n+1)n\,\xi^i\partial_{k_i}\partial_E^{\,n-1}\widetilde{O}(E,\mathbf{k})
\Big].
\label{scalarwardmo}
\end{align}

When boost eigenvalues are continued to imaginary values, we need to replace $\xi_i \to -i\xi_i$ in the Ward identities. 

\section{GCA two-point function}
In this section, we study the two-point function of GCA scalar primaries. First, we write down the Ward identities for two-point functions. Then we solve those Ward identities to obtain the two-point function. Following that, we obtain the momentum space two-point function via Fourier transform from position space. We will perform an independent check by deriving them via the nonrelativistic limit outlined in \ref{psn}. 
\subsection{Two-point function from Ward identities}

We determine the two-point function by imposing the Ward identities of the
finite-dimensional Galilean conformal algebra. Translation invariance implies
that the correlator depends only on the differences of energies and momenta,
which we denote by
\(
G^{(2)}(k,E)
\equiv
\langle \mathcal{O}_1(E_1,k_1)\mathcal{O}_2(E_2,k_2)\rangle
\),
with \(k\equiv k_1=-k_2\) and \(E\equiv E_1=-E_2\).
Under the Galilean boost generator \(M_0\), the Ward identity reads
\begin{align}
\left(-k\,\partial_E+\xi_1+\xi_2\right)G^{(2)}(k,E) &= 0 ,
\end{align}
which integrates to
\begin{align}
G^{(2)}(k,E) &= C(k)\,e^{\frac{\xi}{k}E},
\qquad
\xi\equiv\xi_1+\xi_2 ,
\end{align}
where \(C(k)\) is an arbitrary function of the spatial momentum.
Next, imposing the dilatation Ward identity \(L_0\),
\begin{align}
\left(-E\partial_E-k\partial_k+\Delta_1+\Delta_2-2\right)
C(k)e^{\frac{\xi}{k}E} &= 0 ,
\end{align}
we obtain
\begin{align}
-\frac{\xi E}{k}C(k)e^{\frac{\xi}{k}E}
-k\frac{dC(k)}{dk}e^{\frac{\xi}{k}E}
+\frac{\xi E}{k}C(k)e^{\frac{\xi}{k}E}
+(\Delta-2)C(k)e^{\frac{\xi}{k}E} &= 0 ,
\end{align}
which simplifies to
\begin{align}
-k\frac{dC(k)}{dk}+(\Delta-2)C(k) &= 0 ,
\qquad
\Delta\equiv\Delta_1+\Delta_2 .
\end{align}
Integrating this equation gives
\begin{align}
\frac{dC}{C} &= (\Delta-2)\frac{dk}{k},
\qquad
\Rightarrow\qquad
C(k)=C^{(2)}k^{\Delta-2}.
\end{align}
At this stage, two Ward identities remain. The second boost Ward identity \(M_1\)
takes the form
\begin{align}
\left(-k\,\partial_E^2+2\xi_1\,\partial_E\right)G^{(2)}(k,E) &= 0 ,
\end{align}
which, upon substitution of the explicit solution, becomes
\begin{align}
\left[-\xi_2\,\partial_E G^{(2)}(k,E)+\xi_1\,\partial_E G^{(2)}(k,E)\right] &= 0 ,
\end{align}
implying
\begin{align}
\xi_1=\xi_2 .
\end{align}
Finally, the temporal special conformal Ward identity \(L_1\) gives
\begin{align}
\Big[
-E\partial_E^2
-2k\,\partial_k\partial_E
-2\big((2-\Delta_1)\partial_E-\xi_1\partial_k\big)
\Big]
G^{(2)}(k,E) &= 0 .
\end{align}
Substituting the explicit form of the correlator yields
\begin{align}
-2\left(\Delta-2\Delta_1\right)\xi_1
k^{\Delta-3}e^{\frac{2\xi_1}{k}E} &= 0 ,
\end{align}
which requires
\begin{align}
\Delta_1=\Delta_2 .
\end{align}
Collecting all constraints, the final form of the non-vanishing two-point
function is
\begin{align}
G^{(2)}(k,E)
&= C^{(2)}k^{2\Delta_1-2}
\exp\!\left(\frac{2\xi_1}{k}E\right)=C^{(2)}k^{\Delta-2}
\exp\!\left(\frac{\xi}{k}E\right),
\end{align}
with the selection rules
\begin{align}
\Delta_1=\Delta_2,
\qquad
\xi_1=\xi_2 .
\end{align}

Thus, we obtain a well-defined two-point function by solving the Ward identities. However, just like the position space two-point function, this too diverges as $E \to 0$ and does not admit a Fourier transform. To perform a Fourier transform, it is necessary to analytically continue to a regime where $\xi \to -i\xi$. In this regime, the two-point function becomes:
\begin{align}
G^{(2)}(k,E)
&= C^{(2)}k^{2\Delta_1-2}
\exp\!\left(\frac{2\xi_1}{k}E\right)=C^{(2)}k^{\Delta-2}
\exp\!\left(\frac{-i\xi}{k}E\right),
\end{align}

\subsection{Two-point function via Fourier transform}
The position space correlator was calculated to be:
\[
G^{(2)} =  C\tau^{-\Delta}\,
e^{\frac{\xi x}{\tau}}.
\]
As noted before, this two-point function diverges with distance, and hence its Fourier transform does not exist for real momenta. To obtain a Fourier transform we can either work with imaginary momenta or continue the boost eigenvalues to be imaginary. We choose the latter route:
$\xi \to -i\xi,\,\, G^{(2)} \to  C\tau^{-\Delta}\,
e^{\frac{-i\xi x}{\tau}}$

Now we can perform the Fourier transform in two steps: 

\begin{align}
\notag
C\int_{-\infty}^{\infty} d\tau\,\tau^{-\Delta}e^{i E\tau}\,
\int_{-\infty}^{\infty}\,dx\, e^{-i k x}\, e^{\frac{-i \xi x}{\tau}}
&=C\int_{-\infty}^{\infty} d\tau\,\tau^{-\Delta}e^{i E\tau} \int_{-\infty}^{\infty} d x\,
e^{-i\left({k}+\frac{{\xi}}{\tau}\right){x}} \\ \notag
= C\int_{-\infty}^{\infty} d\tau\,\tau^{-\Delta}e^{i E\tau}\,(2\pi)^d\, \delta\!\left(
k+\frac{\xi}{\tau}
\right)
&=C|\xi|^{1-\Delta} {k}^{\Delta-2}\,
e^{-i\frac{\xi E}{{k}}}=C^{(2)}{k}^{\Delta-2}\,
e^{-i\frac{\xi E}{{k}}}
\end{align}
We thus have: 
\begin{equation}
\label{ft2mom}
  G^{(2)}(E,{k})
\;\propto  {k}^{\Delta-2}\,
e^{-i\frac{\xi E}{{k}}}.
\end{equation}
This is the same result we obtained from solving the Ward identities.
\subsection{Two-point function from non-relativistic limit}

We will now derive the two-point function in a third way, by taking the nonrelativistic limit outlined in Section \ref{psn}.

In a 2D CFT, the two-point function in momentum space is given by
\[
G_2 \;\sim\; (k_+)^{2h-1}(k_-)^{2 \bar{h}-1},
\]
where $k_\pm=k \pm E$ and $\Delta=h+\bar{h}$

As in position space, we take the limit as:
\[
h = \frac{\Delta}{2} + \frac{\xi}{2\gamma},
\qquad
\bar{h} = \frac{\Delta}{2} - \frac{\xi}{2\gamma}.
\]

Then
\[
2h-1 = \Delta - 1 + \frac{2\xi}{\gamma},
\qquad
2\bar h - 1 = \Delta - 1 - \frac{2\xi}{\gamma}.
\]

We write the lightcone momenta as
\[
k_+ = k + \gamma E,
\qquad
k_- = k - \gamma E.
\]

Thus,
\begin{align}
G^{(2)}
&\sim (k_+)^{2h-1}(k_-)^{2\bar h-1} \\
&= (k_+ k_-)^{\Delta-1}\,
(k_+)^{\frac{\xi}{\gamma}}\,
(k_-)^{-\frac{\xi}{\gamma}} \\
\end{align}

Following the same steps as for position space, we obtain in the $\gamma \to 0$ limit: 
\[
G^{(2)}(k,E) \;=\; k^{\Delta-2}\,
\exp\!\left(\frac{\xi E}{k}\right).
\]
up to a constant.
If we started with imaginary boost eigenvalues $\xi \to -i \xi$, we would have obtained (up to an overall constant):
\[
G^{(2)} \;=\; k^{\Delta-2}\,
\exp\!\left(\frac{-i\xi E}{k}\right).
\]
This agrees with our previous result. 

Thus, three different ways of computing the momentum space two-point function converge on the same result.

\section{GCA three-point function} 
In this section, we will study the three-point function of GCA scalar primaries. As with the two-point functions, we will start by writing down the Ward identities. Then we will obtain the momentum space three-point function via Fourier transform from the position space three-point function. This three-point function turns out to satisfy all the Ward identities, establishing it as a valid momentum space three-point function.
 We will then proceed to solve the Ward identities in full generality and obtain an expression for the three-point function. This expression turns out to match exactly with the  Fourier-transformed three-point function, up to a constant. 
\subsection{Ward identities for three-point function}
The Ward identities for the three-point functions are obtained in the same way as the Ward identities for the two-point functions. We start by writing the three-point function as:
\[ G^{(3)}=\delta(k_1+k_2+k_3)\delta(E_1+E_2+E_3)G'^{(3)}\]
where $G'^{(3)}$ is the stripped three-point function. Acting on the three-point function with the generators, one obtains the following Ward identities for the stripped three-point function:
\begin{alignat}{3}
\label{w31}
M_0:\quad&\Bigl[k_1\,\partial_{E_1}+ k_2\,\partial_{E_2}- \xi_1 - \xi_2 - \xi_3\Bigr]\,G'^{(3)}
&\;=\;& 0 \,,\\[6pt]
\label{w32}
M_1:\quad&\Bigl[- k_{1}\,\partial_{E_1}^2+ 2 \xi_1\,\partial_{E_1}- k_{2}\,\partial_{E_2}^2+ 2 \xi_2\,\partial_{E_2}\Bigr]\,G'^{(3)}&\;=\;& 0 \,,\\[6pt]
\label{w33}
L_0:\quad&\Biggl[\sum_{b=1}^{2}\Bigl(- E_b\,\partial_{E_b}- k_b\,\partial_{k_b}\Bigr)- 2(d+1)
+ \Delta_1 + \Delta_2 + \Delta_3\Biggr]\,G'^{(3)}&\;=\;& 0 \,,\\[6pt]
L_1:\quad&\Biggl[- E_1\,\partial_{E_1}^2+ 2\Bigl(- k_1\,\partial_{k_1}- d - 1 + \Delta_1
\Bigr)\partial_{E_1}+ 2 \xi_1\,\partial_{k_1}\notag\\
&\qquad- E_2\,\partial_{E_2}^2+ 2\Bigl(- k_2\,\partial_{k_2}- d - 1 + \Delta_2\Bigr)\partial_{E_2}
+ 2 \xi_2\,\partial_{k_2}\Biggr]\,G'^{(3)}&\;=\;& 0 \,.\label{w34}
\end{alignat}

The derivations are straightforward, except for $M_1$ and $L_1$, which involve the use of other Ward identities. The derivations of these two Ward identities are given in the Appendix \ref{ward}.

\subsection{Three-point function via Fourier transform} 
We start from the position space three-point function \eqref{3p}  and take its Fourier transform: 

\begin{align}
\notag
{G'}^{(3)}(E_i,k_i)
=
\int dt_1\,dt_2\,dx_1\,dx_2\;&
\exp\!\left[
i\left(E_1 t_1 - k_1 x_1 + E_2 t_2 - k_2 x_2\right)
\right] \\ &
\frac{
\exp\!\left[
-i\left(
\xi_{12,3}\frac{x_1-x_2}{t_1-t_2}
+ \xi_{23,1}\frac{x_2}{t_2}
+ \xi_{31,2}\frac{x_1}{t_1}
\right)
\right]
}{
|t_1-t_2|^{\Delta_{12,3}}
|t_2|^{\Delta_{23,1}}
|t_1|^{\Delta_{31,2}}
}
\end{align}

Note that this is the stripped three-point function because, as we saw in \ref{cor}, we have already used translational invariance to get this form of the position space three-point function. 

Doing the x-integrals first:

\begin{align}
\notag \int dx_1\,dx_2\;
\exp\!\Bigg[
-i\Big[
\Big(k_1 &+ \frac{\xi_{12,3}}{t_1-t_2} + \frac{\xi_{31,2}}{t_1}\Big)x_1
+ \Big(k_2 - \frac{\xi_{12,3}}{t_1-t_2} + \frac{\xi_{23,1}}{t_2}\Big)x_2
\Big]
\Bigg]\\
&= \delta\!\left(
k_1 + \frac{\xi_{12,3}}{t_1-t_2} + \frac{\xi_{31,2}}{t_1}
\right)
\delta\!\left(
k_2 - \frac{\xi_{12,3}}{t_1-t_2} + \frac{\xi_{23,1}}{t_2}
\right).
\end{align}

Hence,
\begin{align}
{G'}^{(3)}
&=
\int dt_1\,dt_2\;
e^{i(E_1 t_1 + E_2 t_2)}
|t_1-t_2|^{-\Delta_{12,3}}
|t_2|^{-\Delta_{23,1}}
|t_1|^{-\Delta_{31,2}} \nonumber\\
&\qquad \times
\delta\!\left(
k_1 + \frac{\xi_{12,3}}{t_1-t_2} + \frac{\xi_{31,2}}{t_1}
\right)
\delta\!\left(
k_2 - \frac{\xi_{12,3}}{t_1-t_2} + \frac{\xi_{23,1}}{t_2}
\right).
\end{align}

The delta functions impose the constraints: 
\[
A_1=k_1 + \frac{\xi_{12,3}}{t_1-t_2} + \frac{\xi_{31,2}}{t_1} = 0.
\]

\[A_2=
k_2 - \frac{\xi_{12,3}}{t_1-t_2} + \frac{\xi_{23,1}}{t_2} = 0.
\]

Solving $A_1=0,A_2=0$ simultaneously:

\begin{align}
\label{t1}t_1^{\pm}
&=
\frac{
- \xi_{12,3}(k_1+k_2)
- \xi_{23,1}k_1
- \xi_{31,2}(2k_1+k_2)
\pm \sqrt{\Delta}
}{
2k_1(k_1+k_2)
},\\
\label{t2}
t_2^{\pm}
&=
\frac{
- \xi_{12,3}(k_1+k_2)
- \xi_{23,1}(k_1+2k_2)
- \xi_{31,2}k_2
\pm \sqrt{\Delta}
}{
2k_2(k_1+k_2)
}.
\end{align}

with
\begin{align}
\Delta &=
\xi_{12,3}^2 (k_1+k_2)^2
+ \xi_{23,1}^2 k_1^2
+ \xi_{31,2}^2 k_2^2 \nonumber\\
&\quad
+ 2\xi_{12,3}\xi_{23,1}k_1(k_1+k_2)
+ 2\xi_{12,3}\xi_{31,2}k_2(k_1+k_2)
- 2\xi_{23,1}\xi_{31,2}k_1 k_2 .
\end{align}

The final expression is given by:

\begin{align}
\label{ftf}
G'^{(3)}(E_i,k_i)
&=
\sum_{{(t_1,t_2)=(t_1^\pm,t_2^\mp)}}
\frac{
\exp\!\big[i(E_1 t_1 + E_2 t_2)\big]\;
|t_{12}|^{-\Delta_{12,3}}
|t_2|^{-\Delta_{23,1}}
|t_1|^{-\Delta_{31,2}}
}{
|\det J(t_1,t_2)|
}.
\end{align}

where the Jacobian is given by: 
\[
\det J(t_1,t_2) =\operatorname{det}\left(\frac{\partial (A_1,A_2)}{\partial(t_1,t_2)}\right)
=
\frac{\xi_{12,3}\xi_{23,1}}{t_2^2(t_1-t_2)^2}
+
\frac{\xi_{12,3}\xi_{31,2}}{t_1^2(t_1-t_2)^2}
+
\frac{\xi_{23,1}\xi_{31,2}}{t_1^2 t_2^2}.
\]

Previously, we had seen that the  Fourier-transformed two-point function matches with the solution of the Ward identities.
Does the  Fourier-transformed three-point function also solve the full set of Ward identities?

One of our key results is that it does. The check involves acting on \eqref{ftf} with the Ward identities \cref{w31,w32,w33,w34}. The checks are lengthy and given in the Appendix \ref{proof}. The result establishes the consistency between Fourier transform from position space and momentum space Ward identities for the three-point function. In the next section, we will demonstrate that the  Fourier-transformed three-point function is, in fact, the most general solution to the Ward identities.

\subsection{Solution of Ward identity and matching with  Fourier-transformed three-point function}
In relativistic conformal field theories, a standard strategy for solving Ward identities is to identify quantities that are invariant along the characteristic flows generated by the symmetry operators \cite{Coriano:2013jba,Gupta:2024yib}. Accordingly, we perform a change of variables from \((E_1, E_2, k_1, k_2)\) to a set of dimensionless variables \((x,y,r)\), keeping \(k_1\) as the overall scale.

From the dilatation Ward identity, it follows that the correlator must be a homogeneous function of its arguments. This allows us to factor out the overall scaling dependence as
\begin{equation}
G(E_1,E_2;k_1,k_2)
= k_1^{\,h}\,F(x,y,r),
\qquad
h=\Delta_1+\Delta_2+\Delta_3-2(d+1),
\end{equation}
where we have introduced the dimensionless invariants
\begin{equation}
x=\frac{E_1}{k_1},
\qquad
y=\frac{E_2}{k_1},
\qquad
r=\frac{k_2}{k_1}.
\end{equation}

In terms of these variables, the derivatives appearing in the Ward identities transform as
\begin{equation}
\partial_{E_1}=\frac{1}{k_1}\,\partial_x,
\qquad
\partial_{E_2}=\frac{1}{k_1}\,\partial_y,
\qquad
\partial_{k_1}\;\to\;
\partial_{k_1}
-\frac{x}{k_1}\,\partial_x
-\frac{y}{k_1}\,\partial_y
-\frac{r}{k_1}\,\partial_r .
\end{equation}

With this parametrization, the dilatation Ward identity \(L_0\) is satisfied identically. 
In these variables, it takes the form
\begin{align}
-\left(
k_1\partial_{k_1}+k_2\partial_{k_2}
+E_1\partial_{E_1}+E_2\partial_{E_2}
\right)G
+h\,G &= 0\implies k_1\partial_{k_1}G=hG
\end{align}
which is automatically satisfied by the homogeneous ansatz
\(G=k_1^{\,h}F(x,y,r)\).
Consequently, the remaining Ward identities reduce to differential
constraints on the reduced function \(F(x,y,r)\).

In terms of the variables \((x,y,r)\), the boost Ward identity \(M_0\) becomes
\begin{equation}
\left(
k_1\partial_{E_1}+k_2\partial_{E_2}+i\Xi
\right)F = 0 ,
\qquad
\Xi \equiv \xi_1+\xi_2+\xi_3 .
\end{equation}
Using \(\partial_{E_1}=k_1^{-1}\partial_x\) and
\(\partial_{E_2}=k_1^{-1}\partial_y\), this equation reads
\begin{equation}
\left(\partial_x+r\,\partial_y+i\Xi\right)F=0 .
\end{equation}

To decouple the energy derivatives, we introduce the linear combinations
\begin{equation}
u = x+r\,y,
\qquad
v = r\,x-y,
\end{equation}
for which
\begin{equation}
\partial_x=\partial_u+r\,\partial_v,
\qquad
\partial_y=r\,\partial_u-\partial_v .
\end{equation}
The \(M_0\) Ward identity then reduces to
\begin{equation}
(1+r^2)\,\partial_u F = -i\Xi\,F ,
\end{equation}
whose solution is
\begin{equation}
F(u,v,r)=e^{-i\alpha(r)\,u}\,\Phi(v,r),
\qquad
\alpha(r)=\frac{\Xi}{1+r^2}.
\end{equation}

The second boost Ward identity \(M_1\) takes the form
\begin{equation}
\left(\partial_x^2+r\,\partial_y^2+i2\xi_1\partial_x+i2\xi_2\partial_y\right)F = 0 .
\end{equation}
Substituting \(F=e^{-i\alpha u}\Phi\) and expressing derivatives in
\((u,v)\) coordinates, this equation becomes
\begin{equation}
-(r^2+r)\,\Phi_{vv}(v,r)+iB(r)\,\Phi_v(v,r)+C(r)\,\Phi(v,r)=0 ,
\end{equation}
where
\begin{align}
B(r) &= 2\alpha r(1-r)-2(\xi_1 r-\xi_2), \\
C(r) &= \alpha^2(1+r^3)-2\alpha (\xi_1+r\,\xi_2).
\end{align}

The general solution is therefore
\begin{equation}
\Phi(v,r)
= A(r)\,e^{-i\lambda_+(r)\,v}
+ B(r)\,e^{-i\lambda_-(r)\,v},
\end{equation}
where we inserted $-i$ for later convenience and \(\lambda_\pm(r)\) are the roots of
\begin{equation}
(r^2+r)\lambda^2+B(r)\lambda+C(r)=0 .
\end{equation}
It is given by 
\begin{equation}
    \lambda_{\pm}=\frac{-(1-r)(1+r)^2\xi_{12,3}-r\xi_{13,2}+r^2(2+r)\xi_{13,2}-(1+2r-r^2)\xi_{23,1}\pm(1+r^2)\sqrt{\Delta_{\xi}}}{2r(1+r)(1+r^2)}
\end{equation}
with 
$$\Delta_{\xi}=(1+r)^2\xi_{12,3}^2+(\xi_{23,1}-r\xi_{13,2})^2+2\xi_{12,3}(1+r)(r\xi_{13,2}+\xi_{23,1})$$

The remainder $A(r)$ and $B(r)$ are determined via $L_1$ ward identity which in these coordinates becomes:
\begin{multline}
    \bigg[-x \partial_x^2- y\partial_y^2+2\big(-k_1\partial_{k_1}+x\partial_{x}+y\partial_{y}+r\partial_{r}-d+\Delta_1\big)\partial_x	-2i\xi_1k_1\partial_{k_1} +2i\xi_1(x\partial_{x}+y\partial_{y}+r\partial_{r})\\
    + 2\big(-r\partial_{r}-d-1+\Delta_2\big)\partial_y- 2i\xi_2\partial_{r}\bigg]G=0
\end{multline}
This fixes $A(r)$ up to some constant $\mathbf{c}_1$ as:
\begin{equation}
    A(r) = \mathbf{c}_1\frac{ (r+1)^{2(\Delta_3-1)}
a^{\,1-\Delta_1}b^{\,\Delta_2-1}c^{\,1-\Delta_3}}{\sqrt{\mathcal{D}}}\,.
\end{equation}
where
\begin{align}
a &=
\xi_{12,3}
\big(
\xi_{13,2}+\xi_{23,1}+\sqrt{\mathcal{D}}+2\xi_{13,2} r
\big)
+ \xi_{13,2}
\big(
-\xi_{23,1}+\sqrt{\mathcal{D}}+\xi_{13,2} r
\big)
+ \xi_{12,3}^2 (r+1), \\[4pt]
b &=
\xi_{12,3}
\big(
2\xi_{23,1}+\sqrt{\mathcal{D}}+r(\xi_{13,2}+\xi_{23,1})
\big)
+ \xi_{23,1}
\big(
\xi_{23,1}+\sqrt{\mathcal{D}}-\xi_{13,2} r
\big)
+ \xi_{12,3}^2 (r+1), \\[4pt]
c &=
(\xi_{13,2}+\xi_{23,1})
\big(
\xi_{23,1}+\sqrt{\mathcal{D}}-\xi_{13,2} r
\big)
- \xi_{12,3} (r+1)(\xi_{13,2}-\xi_{23,3}),\\[4pt]
\mathcal{D}&=  (1+r)^2 \xi_{12,3}^2+ 2(1+r)\xi_{12,3}(r\xi_{13,2}+\xi_{23,1})+ (\xi_{23,1}-r\xi_{13,2})^2 .
\end{align}
and similarly, $B(r)$ is given as:
\begin{equation}
B(r)=\mathbf{c}_2\frac{\,r^{2(\Delta_2-1)} a^{\,\Delta_3-1}b^{\,\Delta_1-1} c^{\,1-\Delta_2}}{\sqrt{\mathcal{D}}}\,.
\end{equation}

This three-point function matches exactly with the Fourier-transformed three-point function derived in the previous section. To match with \eqref{t1} and \eqref{t2}, we note that the coefficient of $iE_1$ and $iE_2$ are given as:
\begin{align}
    -\frac{\alpha+r\lambda_{\pm}}{k_1} &=\frac{
- \xi_{12,3}(k_1+k_2)
- \xi_{23,1}k_1
- \xi_{31,2}(2k_1+k_2)
\mp \sqrt{\Delta}
}{
2k_1(k_1+k_2)
}=t_{1}^\mp,\\
    -\frac{\alpha r-\lambda_{\pm}}{k_1}&=\frac{
- \xi_{12,3}(k_1+k_2)
- \xi_{23,1}(k_1+2k_2)
- \xi_{31,2}k_2
\pm \sqrt{\Delta}
}{
2k_2(k_1+k_2)
}=t_{2}^\pm.
\end{align}

The $r$ dependence of the $A(r)$ and $B(r)$ are identical to their fourier transformed counterpart upto some constant and it was confirmed by taking the $r$ derivative of the ratio of Fourier transformed three point function to above. The derivative is found to vanish and this check confirms that the two functions agree. Further, it allows us to extract the proportionality constant. To match the above three-point function with  Fourier-transformed version, we need $\mathbf{c}_1$ and $\mathbf{c}_2$ to depend on the constants involved as follows:
$$\mathbf{c}_1=C_12^{\,2+2\Delta_3}\,
\sqrt{\mathcal{D}_1}\,
K_1^{\,\Delta_3-1}
K_2^{\,\Delta_1-1}
K_3^{\,1-\Delta_2}
L_1^{\,-\Delta_1+\Delta_2-\Delta_3}
L_2^{\,\Delta_1-\Delta_2-\Delta_3}
L_3^{\,-\Delta_1-\Delta_2+\Delta_3}
\,\mathcal{N}$$
$$\mathbf c_2=C_2
2^{4\Delta_3}\,
\sqrt{\mathcal{D}_1}\,
K_1^{\,1-\Delta_1}
K_2^{\,\Delta_2-1}
K_3^{\,1-\Delta_3}
L_1^{\,\Delta_1-\Delta_2-\Delta_3}
L_2^{\,-\Delta_1+\Delta_2-\Delta_3}
L_3^{\,-\Delta_1-\Delta_2+\Delta_3}
\,\mathcal{N}$$
where,
\begin{align*}
\mathcal{D}_1 &=
4\xi_{12,3}^2
+ 4\xi_{12,3}(\xi_{13,2}+\xi_{23,1})
+ (\xi_{23,1}-\xi_{13,2})^2 , \\[4pt]
\mathcal{D}_2 &=
2\xi_{23,1}(2\xi_{12,3}-\xi_{13,2})
+ (2\xi_{12,3}+\xi_{13,2})^2
+ \xi_{23,1}^2 , \\[6pt]
K_1 &=
\xi_{12,3}\big(\sqrt{\mathcal{D}_1}+3\xi_{13,2}+\xi_{23,1}\big)
+ \xi_{13,2}\big(\sqrt{\mathcal{D}_1}+\xi_{13,2}-\xi_{23,1}\big)
+ 2\xi_{12,3}^2 , \\[4pt]
K_2 &=
\xi_{12,3}\big(\sqrt{\mathcal{D}_1}+\xi_{13,2}+3\xi_{23,1}\big)
+ \xi_{23,1}\big(\sqrt{\mathcal{D}_1}-\xi_{13,2}+\xi_{23,1}\big)
+ 2\xi_{12,3}^2 , \\[4pt]
K_3 &=
(\xi_{13,2}+\xi_{23,1})
\big(\sqrt{\mathcal{D}_1}-\xi_{13,2}+\xi_{23,1}\big)
- 2\xi_{12,3}(\xi_{13,2}-\xi_{23,1}) , \\[6pt]
L_1 &=
-\sqrt{\mathcal{D}_2}-2\xi_{12,3}-\xi_{13,2}-3\xi_{23,1}, \\[2pt]
L_2 &=
\sqrt{\mathcal{D}_2}-2\xi_{12,3}-3\xi_{13,2}-\xi_{23,1}, \\[2pt]
L_3 &=
\frac{1}{4}\big(\sqrt{\mathcal{D}_2}-2\xi_{12,3}-3\xi_{13,2}-\xi_{23,1}\big)
+ \frac{1}{4}\big(\sqrt{\mathcal{D}_2}+2\xi_{12,3}+\xi_{13,2}+3\xi_{23,1}\big), \\[6pt]
\mathcal{N}^{-1} &=
\frac{256\xi_{13,2}\xi_{23,1}}{L_1^2 L_2^2}
+ \frac{16\xi_{12,3}\xi_{13,2}}{L_1^2 L_3^2}
+ \frac{16\xi_{12,3}\xi_{23,1}}{L_2^2 L_3^2}
\end{align*}

\section{Summary and Future Plans}
In this paper, we have taken the first step towards studying GCA using momentum space methods. By solving the Ward identities, we found the momentum space two-point and three-point functions. In the regime where the boost eigenvalues are continued to imaginary values, these were found to match the results from Fourier transform of position-space correlators. The two-point function was also verified via a nonrelativistic limit of the conformal two-point function. 

Several extensions suggest themselves. First, it would be natural to generalize our momentum-space Ward identity analysis beyond scalar primaries to include operators with spin and conserved currents, in particular the stress-tensor sector. Second, the present methods should extend to four-point functions, providing a momentum-space bootstrap framework for GCA in which factorization channels and singularity structure are manifest. 

Third, since the momentum-space GCA three-point function is significantly simpler than the relativistic conformal counterpart (expressible in terms of triple-K integrals), it would be interesting to derive our result directly as a controlled nonrelativistic limit of the momentum-space CFT correlator. Finally, since BMS$_3$ is isomorphic to GCA$_2$ and momentum space correlators in the boundary are typically closely related to scattering amplitudes in the bulk, our work might help develop a tool to study bulk scattering in 3 dimensions. 

\acknowledgments{
AC and NK would like to thank Arundhati Goldar for valuable discussions. }

\appendix
\section{Derivation of M$_1$ and L$_1$ Ward identities for three-point functions}\label{ward}
We provide the derivations of $M_1$ and $L_1$ Ward identities for the three-point function. The other derivations follow similarly. 
\subsection{ M$_1$ Ward identity for the 3-pt function}

The M$_1$ Ward identity is given by:
\[
\mathcal M_1=\sum_{a=1}^3\left(-k_a\,\partial_{E_a}^2+2\xi_a\,\partial_{E_a}\right).
\]
We define new variables
\[
S := E_1+E_2+E_3,\qquad e_1 := E_1,\qquad e_2 := E_2,
\]
so that
\[
E_3 = S-e_1-e_2.
\]
In these variables, the full three-point function becomes
\[
G^{(3)}(E_1,k_1,E_2,k_2,E_3,k_3)
= \delta(S)\,\delta(K)\,G'(S,e_1,e_2,k_1,k_2,k_3),
\qquad K:=k_1+k_2+k_3.
\]
We will suppress the spectator factor $\delta(K)$ until the end, since the $M_1$ generator differentiates only with respect to energies.

In our new variables, we have
\[
\partial_{E_1}=\partial_S+\partial_{e_1},\qquad \partial_{E_2}=\partial_S+\partial_{e_2},\qquad \partial_{E_3}=\partial_S.
\]
 
We use the notation:
\[
G_S:=\partial_S G',\quad G_1:=\partial_{e_1}G',\quad G_2:=\partial_{e_2}G',
\]
\[
G_{SS}:=\partial_S^2 G',\quad G_{S1}:=\partial_S\partial_{e_1}G',\quad 
G_{S2}:=\partial_S\partial_{e_2}G',
\]
\[
G_{11}:=\partial_{e_1}^2G',\quad G_{22}:=\partial_{e_2}^2G',\quad 
G_{12}:=\partial_{e_1}\partial_{e_2}G'.
\]
Also let $\delta:=\delta(S)$, $\delta':=\frac{d}{dS}\delta(S)$, and $\delta'':=\frac{d^2}{dS^2}\delta(S)$.
Define also
\[
\Xi := \xi_1+\xi_2+\xi_3.
\]
Now we proceed to compute the derivatives of $G^{(3)}$:
\begin{align*}
\partial_{E_1}(\delta G')
&=(\partial_S+\partial_{e_1})(\delta G')
=\partial_S(\delta G')+\partial_{e_1}(\delta G')\\
&=(\delta'G'+\delta G_S)+\big[(\partial_{e_1}\delta)\,G'+\delta(\partial_{e_1}G')\big].
\end{align*}
Since $\delta=\delta(S)$ depends only on $S$, we have $\partial_{e_1}\delta=0$, hence
\begin{align}
\label{m11}
\partial_{E_1}(\delta G')=\delta'G'+\delta(G_S+G_1).
\end{align}
By similar computations, we obtain:
\begin{align}
\label{M12}
\partial_{E_2}(\delta G')&=\delta'G'+\delta(G_S+G_2),\\
\label{M13}
\partial_{E_3}(\delta G')=\delta'G'+\delta\,G_S.
\end{align}

The computation of the second derivatives is straightforward but tedious. The results are:
\begin{align}
\label{M111}
\partial_{E_1}^2(\delta G')
&=\delta''G'+2\delta'(G_S+G_1)+\delta(G_{SS}+2G_{S1}+G_{11})\\
\label{M122}
\partial_{E_2}^2(\delta G')
&=\delta''G'+2\delta'(G_S+G_2)+\delta(G_{SS}+2G_{S2}+G_{22}).\\
\label{M133}
\partial_{E_3}^2(\delta G')
&=\delta''G'+2\delta'G_S+\delta G_{SS}.
\end{align}

Using equations \cref{m11,M12,M13,M111,M122,M133} we evaluate
\[
\mathcal M_1(\delta G')
=\sum_{a=1}^3\Big[-k_a\,\partial_{E_a}^2(\delta G')+2\xi_a\,\partial_{E_a}(\delta G')\Big],
\]
and group terms as
\[
\mathcal M_1(\delta G')=\left(\delta''\,A+\delta'\,B+\delta\,C\right)\delta(K).
\]
where
\begin{align*}
A &= -(k_1+k_2+k_3)\,G' = -K\,G'.\\
B &= -2K\,G_S-2k_1G_1-2k_2G_2+2\Xi\,G'
=2\big[-K\,G_S-k_1G_1-k_2G_2+\Xi\,G'\big].\\
C &= -K\,G_{SS}-2k_1G_{S1}-2k_2G_{S2}-k_1G_{11}-k_2G_{22}
+2\Xi\,G_S+2\xi_1G_1+2\xi_2G_2.
\end{align*}
A vanishes by momentum conservation, B vanishes via the $M_0$ Ward identity. 

The first term of C vanishes by momentum conservation. We note that $$-2k_1G_{S1}-2k_2G_{S2}+2\Xi\,G_S=-2\frac{\partial}{\partial S}(k_1G_1+k_2G_2-\Xi G')$$ which vanishes by the $M_0$ identity again. The final answer is then:
\begin{equation}
  \boxed{M_1(\delta(K)\delta(S)G') =  \delta(K)\delta(S)\left( -k_1G_{11}-k_2G_{22}+2\xi_1G_1+2\xi_2G_2\right).}
\end{equation}

\subsection{L$_1$ Ward identity on the 3-pt function}

In this case, we have to deal with momentum derivatives. We define:
\[
K := k_1 + k_2 + k_3,\, \kappa_1=k_1,\,\kappa_2=k_2.
\]
The derivatives are:
\begin{align*}
\partial_{k_1}&=\partial_K+\partial_{\kappa_1}
\\
\partial_{k_2}&=\partial_K+\partial_{\kappa_2}\\
\partial_{k_3}&=\partial_K
\end{align*}
Introduce the notation:
\[
G_{k_a} := \partial_{k_a} G.
\]
We will denote $\frac{\partial\delta(K)}{\partial K}=\delta'(K)$.

For convenience, we define:
\[
H_1 := G_S + G_1,
\qquad
H_2 := G_S + G_2,
\qquad
H_3 := G_S,
\]

We list the results block by block.\\
\textbf{Block 1:}

\begin{align*}
\sum_a -E_a \partial_{E_a}^2 G^{(3)}
= \delta(K)\Big[
\delta''(S)(-SG)
+ \delta'(S)(-2(SG_S + e_1 G_1 + e_2 G_2))\\
+ \delta(S)C^{(1)}(-SG_{SS}
- 2e_1 G_{S1} - e_1 G_{11}
- 2e_2 G_{S2} - e_2 G_{22})
\Big].
\end{align*}

\textbf{Block 2:}
\[
\begin{aligned}
2\sum_a(-k_a\partial_{k_a}-d-1+\Delta_a)\partial_{E_a}G^{(3)}
={}-2\delta'(K)\Big[
K\delta'(S)G'
+\delta(S)(\kappa_1H_1+\kappa_2H_2+(K-\kappa_1-\kappa_2)H_3)
\Big]\\
-2\delta(K)\Big[
\delta'(S)(KG_K+\kappa_1G_{\kappa_1}+\kappa_2G_{\kappa_2})
+\delta(S)\mathcal{H}
\Big]\\
+\delta(K)\delta'(S)\Big(2\sum_a(-d-1+\Delta_a)\Big)G'\\
+\delta(K)\delta(S)\Big(2\sum_a(-d-1+\Delta_a)H_a\Big).
\end{aligned}
\]

where
\[
\mathcal{H}
:=\kappa_1(H_{1,K}+H_{1,\kappa_1})
+\kappa_2(H_{2,K}+H_{2,\kappa_2})
+(K-\kappa_1-\kappa_2)H_{3,K}.
\]
\textbf{Block 3:}

\[
2\sum_a\xi_a\partial_{k_a}G^{(3)}
=\delta(S)\delta'(K)(2\Xi G)
+\delta(S)\delta(K)\big[2\Xi G_K+2\xi_1G_{\kappa_1}+2\xi_2G_{\kappa_2}\big].
\]

Putting the terms together, we get: 

\[ L_1G = A\delta(K)\delta''(S)+B\delta'(K)\delta'(S) +C\delta(K)\delta'(S)+D\delta'(K)\delta(S)+E\delta(K)\delta(S)\]

where 
\begin{align}
A=-SG'
\end{align}
\begin{align}
B=-2KG'
\end{align}
\begin{align}\notag
C&=-2KG_K +2 \left(e_1G_1+e_2G_2-\kappa_1G_{\kappa_1}+\kappa_2G_{\kappa_2}-2\Sigma_a((d+1)+\Delta_a)G'\right)-2SG_S\\
&=-2SG_S
\end{align}
The first term vanished by conservation of momentum, while the second term vanished by the $L_0$ Ward identity. 
\begin{align}
D=-2(  \kappa_1G_{1}+\kappa_2G_{2}-\Xi G')-2KG_S=-2KG_S  
\end{align}
where the first term vanished by the $M_1$ Ward identity. 
From adding B and D,we get:
\begin{equation}
 B+D=-2K \delta'(K) (G'\delta'(S)+G_S \delta(S))=-2K\delta'(K)\frac{\partial}{\partial S}(G'\delta (S))=0.   
\end{equation}
where the last identity is understood to hold distributionally. 
From adding A and C: 
\begin{align}
 &-SG \delta(K)\delta''(S) -2SG_S\delta(K)\delta'(S)=(G'+SG_S-2SG_S)\delta(K)\delta'(S)\\
 &=(G'-SG_S)\delta(K)\delta'(S)=(G_S-SG_{SS}-G_S)\delta(K)\delta(S)=SG_SS\delta(K)\delta(S)=0
\end{align}
where we used integration by parts in the second and fourth steps. The last term vanishes by energy conservation. Again, this is understood to hold distributionally. 

Only the E term survives: 
\begin{align}\notag
E&=-\Big(e_1G_{11}+e_2G_{22}-2\kappa_1G_{\kappa_1}-2\kappa_2G_{\kappa_2}+\sum_a 2(-d-1+\Delta_a)G_a+2\xi_1G_{\kappa_1}+2\xi_2G_{\kappa_2}\Big)\\ \notag
&-SG_{SS}-2KG_{SK}+2\frac{\partial}{\partial S}\Big(e_1G_1+e_2G_2+\kappa_1G_{\kappa_1}+\kappa_2G_{\kappa_2}+\sum_a(-d-1+\delta_a)G'\Big)\\
&-2\frac{\partial}{\partial K}\Big(\kappa_1G_1+\kappa_2G_2-\Xi G')
\end{align}
The second term vanishes by energy conservation and the third term by momentum conservation. The fourth and fifth terms vanish by $L_1$ and $M_1$ Ward identities respectively.

We have the final form of the $L_1$ identity:
\[ \boxed{e_1G_{11}+e_2G_{22}-2\kappa_1G_{\kappa_1}-2\kappa_2G_{\kappa_2}+2\sum_a (-d-1+\Delta_a)G_a+2\xi_1G_{\kappa_1}+2\xi_2G_{\kappa_2}=0.}\]

\section{Proof of the Fourier-transformed three-point function satisfying the Ward identities}\label{proof}

In this section, we will show that the  Fourier-transformed three-point function satisfies the full set of Ward identities. 

We start by recalling the three-point function as a sum over branches.
On each branch $\sigma$,
\begin{equation}
G^{(\sigma)}(E_1,E_2;k_1,k_2)
=
\mathcal N^{(\sigma)}(k;\Delta,\xi)\,
e^{i(E_1 t_1^{(\sigma)}+E_2 t_2^{(\sigma)})},
\end{equation}
where \[
\mathcal{N}^{(\sigma)} =
\frac{|t_1 - t_2|^{-\Delta_{12,3}} |t_2|^{-\Delta_{23,1}} |t_1|^{-\Delta_{31,2}}}
{|\det J|},
\]
with $\mathcal N^{(\sigma)}$ independent of $E_{1,2}$.
The saddle-point variables $(t_1,t_2)$ are fixed by the constraints
\begin{align}
\label{a1} A_1 &:= k_1 + \frac{\alpha}{t_1-t_2} + \frac{\gamma}{t_1} = 0,\\
\label{a2} A_2 &:= k_2 - \frac{\alpha}{t_1-t_2} + \frac{\beta}{t_2} = 0,
\end{align}
with
\begin{equation}
\label{alpha} \alpha = \xi_1+\xi_2-\xi_3,\qquad
\beta = \xi_2+\xi_3-\xi_1,\qquad
\gamma = \xi_3+\xi_1-\xi_2 .
\end{equation}

\subsection{Checking the $M_0$ Ward identity }
The $M_0$ Ward identity is given by:
\begin{equation}
\bigl[k_1\partial_{E_1}+k_2\partial_{E_2}
+i(\xi_1+\xi_2+\xi_3)\bigr]G=0.
\end{equation}
\begin{equation}
\partial_{E_1} G^{(\sigma)} = i t_1 G^{(\sigma)},\qquad
\partial_{E_2} G^{(\sigma)} = i t_2 G^{(\sigma)},\qquad
\partial_{E_1}\partial_{E_2} G^{(\sigma)} = -t_1 t_2 G^{(\sigma)} .
\end{equation}
Acting on $G^{(\sigma)}$,
\begin{equation}
\left(k_1\partial_{E_1}+k_2\partial_{E_2}\right)G
=
i(k_1 t_1+k_2 t_2)G.
\end{equation}
Multiplying \eqref{a1} by $t_1$, \eqref{a2} by $t_2$ and adding the resulting identities, we get:
\begin{equation}
k_1 t_1 + k_2 t_2 + (\alpha+\beta+\gamma) = 0 \implies k_1 t_1 + k_2 t_2 = -(\xi_1+\xi_2+\xi_3).
\end{equation}
Hence
\begin{equation}
i(k_1 t_1+k_2 t_2)+i(\xi_1+\xi_2+\xi_3)
= i\bigl[-(\xi_1+\xi_2+\xi_3)+(\xi_1+\xi_2+\xi_3)\bigr]=0,
\end{equation}
The $M_0$  Ward identity is satisfied identically.

\subsection{Checking $M_1$ Ward identity:}

The $M_1$ Ward identity is given by
\begin{align}
    [-k_{1}\partial_{E_1}^2+2\xi_1\partial_{E_1}-k_{2}\partial_{E_2}^2+2\xi_2\partial_{E_2}]G=0
\end{align}
using $M_0$, it can be re-written as:
\begin{align*}
	[-\partial_{E_1}(k_{1}\partial_{E_1})-\partial_{E_2}(k_{2}\partial_{E_2})+2\xi_1\partial_{E_1}+2\xi_2\partial_{E_2}] G&=0\\
	[-\partial_{E_1}(-k_{2}\partial_{E_2}+\xi_1+\xi_2+\xi_3)-\partial_{E_2}(-k_{1}\partial_{E_1}+\xi_1+\xi_2+\xi_3)+2\xi_1\partial_{E_1}+2\xi_2\partial_{E_2}] G&=0\\
    \Bigl[(k_1+k_2)\partial_{E_1}\partial_{E_2}
    -i(\xi_1-\xi_2-\xi_3)\partial_{E_1}
    -i(-\xi_1+\xi_2-\xi_3)\partial_{E_2}\Bigr]G&=0.
\end{align*}

Acting on $G^{(\sigma)}$,
\begin{align}
(k_1+k_2)\partial_{E_1}\partial_{E_2}
&\to -(k_1+k_2)t_1 t_2,\\
-i(\xi_1-\xi_2-\xi_3)\partial_{E_1}
&\to +(\xi_1-\xi_2-\xi_3)x_1,\\
-i(-\xi_1+\xi_2-\xi_3)\partial_{E_2}
&\to +(-\xi_1+\xi_2-\xi_3)x_2.
\end{align}

Using
\begin{equation}
\xi_1-\xi_2-\xi_3=-\beta,\qquad
-\xi_1+\xi_2-\xi_3=-\gamma,
\end{equation}
the condition becomes
\begin{equation}
-(k_1+k_2)t_1 t_2-\beta t_1-\gamma t_2=0.
\end{equation}

Solving for $\alpha/(t_1-t_2)$ in \eqref{a1} and \eqref{a2} and equating, we get:
\begin{equation}
\frac{\alpha}{t_1-t_2} = -k_1 - \frac{\gamma}{t_1}
= k_2 + \frac{\beta}{t_2}.
\end{equation}
Hence
\begin{equation}
k_1+k_2 = \frac{\gamma}{t_1}-\frac{\beta}{t_2}.
\end{equation}
Multiplying by $t_1 t_2$,
\begin{equation}
(k_1+k_2)t_1 t_2 = -\gamma t_2 - \beta t_1 .
\end{equation}

\begin{equation}
(k_1+k_2)t_1 t_2=-\gamma t_2-\beta t_1,
\end{equation}
So the second Ward identity is also satisfied identically.

\subsection{Checking $L_0$ Ward identity}

We have for each branch, the energy derivative part:
\begin{equation}
\label{edp}
- E_1\partial_{E_1}G^{(\sigma)}
- E_2\partial_{E_2}G^{(\sigma)}
= -i(E_1 t_1+E_2 t_2)G^{(\sigma)}.
\end{equation}

and the momentum derivative part:

\begin{equation}
\label{pdp}
-\sum_{b=1}^2 k_b\partial_{k_b}G^{(\sigma)}
=
-(k_1\partial_{k_1}+k_2\partial_{k_2})\ln\mathcal N^{(\sigma)}\,G^{(\sigma)}
-i(E_1\Sigma_1+E_2\Sigma_2)G^{(\sigma)},
\end{equation}
where
\begin{equation}
\Sigma_a := k_1\partial_{k_1}t_a+k_2\partial_{k_2}t_a .
\end{equation}

Now it can be directly checked from eqs \eqref{t1} and \eqref{t2} that\footnote{An easy way to see this is that $[t_1]=[t_2]=-1.$ in $k$.Under $k_i\to\lambda k_i$, $t_1\to\lambda^{-1}t_1,t_2\to\lambda^{-1}t_2.$}:

\begin{equation}
k_1\partial_{k_1}t_a+k_2\partial_{k_2}t_a=-t_a,\qquad a=1,2.
\end{equation}
Therefore
\begin{equation}
-i(E_1\Sigma_1+E_2\Sigma_2)
=-i\bigl[E_1(-t_1)+E_2(-t_2)\bigr]
=+i(E_1 t_1+E_2 t_2).
\end{equation}

This exactly cancels the contribution from \eqref{edp}.

After the cancellation, the $L_0$ Ward identity reduces to:
\begin{equation}
\label{redl1}
(k_1 \partial_{k_1} + k_2 \partial_{k_2}) \mathcal{N}^{(\sigma)}
= (\Delta_1 + \Delta_2 + \Delta_3 - 2(d+1)) \mathcal{N}^{(\sigma)}.
\end{equation}

From the explicit expression,
\[
\mathcal{N}^{(\sigma)} \sim
\frac{|t_1 - t_2|^{-\Delta_{12,3}} |t_2|^{-\Delta_{23,1}} |t_1|^{-\Delta_{31,2}}}
{|\det J|},
\]
with
\[
|\det J| \sim \frac{1}{t_1^2 t_2^2 (t_1 - t_2)^2},
\]
we see that $\mathcal{N}^{(\sigma)}$ is a homogeneous function of degree
\[
\deg(\mathcal{N}^{(\sigma)})
= (\Delta_{12,3} + \Delta_{23,1} + \Delta_{31,2}) - 2\times 2
= (\Delta_1 + \Delta_2 + \Delta_3) - 4.
\]

Thus the LHS of \eqref{redl1} becomes
\[
(k_1 \partial_{k_1} + k_2 \partial_{k_2}) \mathcal{N}^{(\sigma)}
= (\Delta_1 + \Delta_2 + \Delta_3 - 4) \mathcal{N}^{(\sigma)},
\]
After substituting $d=1$ we see that this is exactly equal to the RHS. Hence the $L_0$ Ward identity is found to be satisfied.

\subsection{Checking the $L_1$ Ward identity}

Acting $L_1$ on the  Fourier-transformed three-point function, and grouping by powers of $E_1,E_2$, we get
\[
L_1G = \left(E_1 C_1 + E_2 C_2 + C_0\right)G
\]

where

\[
C_1
= t_1^2
+ 2k_1 t_1 t_{1,1}
+ 2k_2 t_2 t_{1,2}
+ 2\xi_1 t_{1,1}
+ 2\xi_2 t_{1,2}.
\]

\[
C_2
= t_2^2
+ 2k_1 t_1 t_{2,1}
+ 2k_2 t_2 t_{2,2}
+ 2\xi_1 t_{2,1}
+ 2\xi_2 t_{2,2}.
\]

\[
C_0
= 2i\Big[
- k_1 t_{1,1}
- k_1 t_1 L_1
+ (-d - 1 + \Delta_1) t_1
- k_2 t_{2,2}
- k_2 t_2 L_2
+ (-d - 1 + \Delta_2) t_2
\Big]
- 2i(\xi_1 L_1 + \xi_2 L_2).
\]

Here \[
L_j := \partial_{k_j} \ln N,
\qquad
t_{a,j} := \partial_{k_j} t_a
\quad (a=1,2;\; j=1,2).
\]
For $L_1$ to hold for generic $(E_1,E_2)$, we need three separate conditions:
\[
C_1 = 0,
\qquad
C_2 = 0,
\qquad
C_0 = 0.
\]

Now we compute $t_{1,1},t_{1,2},t_{2,1}$ on the saddles.
Differentiating $A_1=A_2=0$ with respect to $k_1,k_2$ gives
\[
J
\begin{pmatrix}
t_{1,1} \\ t_{2,1}
\end{pmatrix}
= -
\begin{pmatrix}
1 \\ 0
\end{pmatrix},
\qquad
J
\begin{pmatrix}
t_{1,2} \\ t_{2,2}
\end{pmatrix}
= -
\begin{pmatrix}
0 \\ 1
\end{pmatrix}.
\]

Solving explicitly using the above $\det J$ gives neat closed forms:
\begin{align}
\label{x1}t_{1,1} &= \frac{t_1^2(\alpha t_2^2 + \beta(t_1-t_2)^2)}{D}\\ \label{x2}
t_{2,2} &= \frac{t_2^2(\alpha t_1^2 + \gamma(t_1-t_2)^2)}{D}\\ \label{x3}
t_{2,1} &= t_{1,2} = \frac{\alpha t_1^2 t_2^2}{D},
\end{align}
where \[
D=\alpha\beta t_1^2+\alpha\gamma t_2^2+\beta\gamma t_{12}^2.
\]

On the constraint surface we can eliminate $k_1,k_2$:
\begin{align}
\label{k1} k_1 &= \frac{\alpha}{t_1-t_2} = \frac{\gamma}{t_1}\\
\label{k2} k_2 &= \frac{\alpha}{t_1-t_2} = \frac{\beta}{t_2}.
\end{align}
\subsubsection*{Checking $C_1=0$}

Using \cref{x1,x3} and doing some algebra, $C_1$ simplifies to 
\[
C_1
= \frac{t_1^2}{D}
\big[
D + 2k_1 t_1 A + 2\alpha k_2 t_2^3 + 2\xi_1 A + 2\alpha \xi_2 t_2^2 \big]
\]
Next we substitute \cref{k1,k2} and with some more algebra we get:
\[
C_1
=\frac{t_1^2}{D}
\Big[
D
-2\alpha(\alpha t_2^2+\beta t_1 t_{12})
-2\gamma(\alpha t_2^2+\beta t_{12}^2)
-2\alpha\beta t_2^2
+2\xi_1(\alpha t_2^2+\beta t_{12}^2)
+2\alpha\xi_2 t_2^2
\Big].
\]

Replace $2\xi_1\to\alpha+\gamma$ and $2\alpha\xi_2\to\alpha(\alpha+\beta)$. Then
\[
C_1
= \frac{1}{D}\Big[D
-2\alpha(\alpha t_2^2+\beta t_1 t_{12})
-2\gamma(\alpha t_2^2+\beta t_{12}^2)
-2\alpha\beta t_2^2
+(\alpha+\gamma)(\alpha t_2^2+\beta t_{12}^2)
+\alpha(\alpha+\beta)t_2^2 \Big]
\]

Finally using \eqref{alpha}, we find that all the terms cancel: 
\[C_1=0.\]

\subsubsection*{Checking $C_0=0$}

We start from:
\[
C_0
=2i\Big[
-k_1 t_{1,1}-k_1 t_1 L_1+(-d-1+\Delta_1)t_1
-k_2 t_{2,2}-k_2 t_2 L_2+(-d-1+\Delta_2)t_2
\Big]
-2i(\xi_1 L_1+\xi_2 L_2),
\]
After making the substitutions and doing the algebra, this simplifies to
\begin{align}
\notag C_0
=2i\Big[\Delta_1 t_1 + \Delta_2 t_2 - (d-1)(t_1+t_2)
-\big((k_1t_1+\xi_1)\big(-\Delta_{12,3}\frac{t_{1,1}-t_{2,1}}{t_{12}}-\Delta_{23,1}\frac{t_{2,1}}{t_2}-\Delta_{31,2}\frac{t_{1,1}}{t_1}\big)\\
+(k_2t_2+\xi_2)\big(-\Delta_{12,3}\frac{t_{1,2}-t_{2,2}}{t_{12}}-\Delta_{23,3}\frac{t_{2,2}}{t_2}-\Delta_{31,2}\frac{t_{1,2}}{t_1}\big)\big)
\Big].\label{co}
\end{align}
Now substituting \cref{x1,x2,x3,k1,k2} in \eqref{co}, and using the definition $\Delta_{ij,k} =\Delta_i + \Delta_j -\Delta_k$, we get: 
\[
C_0=2i\Big[\Delta_1 t_1 + \Delta_2 t_2 - (d-1)(t_1+t_2)-\Big(\Delta_1t_1+\Delta_2t_2\Big)\Big].
\]
Finally, substituting $d=1$, we find that $C_0$ vanishes. 

We have proved that the  Fourier-transformed three-point function satisfies $L_1$.

\bibliography{galilean}
\end{document}